\documentstyle[12pt,prd,aps,epsf,axodraw]{revtex}
\begin{document}
\baselineskip=17pt \draft
\title{Implications of
$\bar{\nu}_{e}e^{-}\rightarrow W^{-}\gamma$ for  high-energy
 $\bar{\nu}_{e}$ observation}
\author{H. Athar$^{1,2,}$\footnote{E-mail: athar@phys.cts.nthu.edu.tw} and
Guey-Lin Lin$^{2,}$\footnote{E-mail: glin@cc.nctu.edu.tw}}
\address{$^1$Physics Division, National Center for Theoretical Sciences,
Hsinchu 300, Taiwan \\
  $^2$Institute of Physics, National Chiao Tung University,
Hsinchu 300, Taiwan}
\date{\today}
\maketitle
\begin{abstract}
\tightenlines

Absorption of high-energy $\bar{\nu}_{e}$ over electrons above the
$W$ boson production threshold is reexamined. It is pointed out
that, in the case of photon emissions along the direction of
incident high-energy $\bar{\nu}_{e}$, the kinematically allowed
 average energy carried by the final state hard photon can be
$\leq 1\%$ of the incident $\bar{\nu}_{e}$ energy above the $W$
boson production threshold. The differential energy spectrum for
the final state hard photon is calculated. We also discuss
implications of our results for the prospective search of
high-energy $\bar{\nu}_{e}$ through this final state hard photon.

\end{abstract}


A positive observation of high-energy neutrinos above the
atmospheric background will mark the beginning of high-energy
neutrino astronomy. Several high-energy neutrino detectors,
commonly known as high-energy neutrino telescopes, are currently
at a rather advanced stage of their deployments. This necessitates
the identification of possible signatures of the high-energy
neutrino interactions, particularly those occurring inside the  high-energy
 neutrino
telescopes. Needless to mention, such studies will complement the high-energy
gamma ray astronomy for understanding the origin of high-energy radiation
from the cosmos \cite{Learned:2000sw}.

In this {\em Brief Report}, we reexamine in some detail the energy
spectrum of the final-state hard photon in the absorption process
$\bar{\nu}_{e}e^{-}\rightarrow W^{-}\gamma$ above the $W$-boson
production threshold. This absorption process has been discussed
in Ref. \cite{Brown:1979ux}, whereas its closely associated
resonant absorption process $\bar{\nu}_{e}e^{-}\rightarrow W^{-}
(\rightarrow \bar{\nu}_{\mu}\mu^{-} $), now referred to as
``Glashow resonance", occurring at
$E^{res}_{\bar{\nu}_{e}}=M^{2}_{W}/(2m_{e})\sim 6.3\cdot 10^{6}$
GeV, was first pointed out in Ref. \cite{shelly}, and was
subsequently studied in Ref. \cite{resonance}. However, to our
knowledge, the energy spectrum of the final-state photon in
$\bar{\nu}_{e}e^{-}\rightarrow W^{-}\gamma$ has not been studied
before. Here, we shall give an estimate of the kinematically
allowed average energy for the hard photon and calculate the
photon energy spectrum as well as the total cross section of the
absorption process $\bar{\nu}_{e}e^{-}\rightarrow W^{-}\gamma$.

\begin{figure}
\begin{center}
{ \unitlength=1.0 pt \SetScale{1.0}
\SetWidth{0.7}      
\scriptsize    
{} \qquad\allowbreak
\begin{picture}(95,79)(0,0)
\Text(15.0,70.0)[r]{$\bar{\nu}_e$}
\ArrowLine(58.0,70.0)(16.0,70.0) \Text(80.0,70.0)[l]{$W^-$}
\DashArrowLine(79.0,70.0)(58.0,70.0){3.0} \Text(54.0,60.0)[r]{$e$}
\ArrowLine(58.0,50.0)(58.0,70.0) \Text(15.0,50.0)[r]{$e$}
\ArrowLine(16.0,50.0)(58.0,50.0) \Text(80.0,50.0)[l]{$\gamma$}
\DashLine(58.0,50.0)(79.0,50.0){3.0} \Text(47,0)[b] {Diagram (a)}
\end{picture} \
{} \qquad\allowbreak
\begin{picture}(95,79)(0,0)
\Text(15.0,70.0)[r]{$\bar{\nu}_e$}
\ArrowLine(37.0,60.0)(16.0,70.0) \Text(15.0,50.0)[r]{$e$}
\ArrowLine(16.0,50.0)(37.0,60.0) \Text(47.0,64.0)[b]{$W^+$}
\DashArrowLine(58.0,60.0)(37.0,60.0){3.0}
\Text(80.0,70.0)[l]{$W^-$}
\DashArrowLine(79.0,70.0)(58.0,60.0){3.0}
\Text(80.0,50.0)[l]{$\gamma$} \DashLine(58.0,60.0)(79.0,50.0){3.0}
\Text(47,0)[b] {Diagram (b)}
\end{picture} \
}
\end{center}
\caption{The Feynman diagrams contributing to $\bar{\nu}_{e}e^{-}\to
W^- \gamma$ in the leading order. \label{fig:feynman}}
\end{figure}
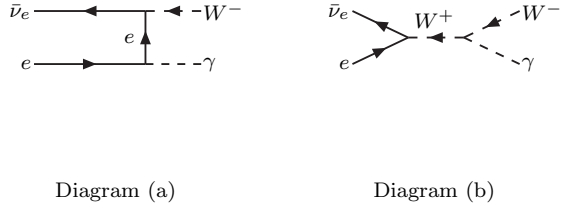

There are two Feynman diagrams contributing to the absorption
process, $\bar{\nu}_{e}e^{-}\rightarrow W^{-}\gamma$  in the
leading order. They are shown in Fig. \ref{fig:feynman}. Out of
these two diagrams,  diagram (b) contains a $W$-boson exchange,
hence its contribution is suppressed for the range of $\sqrt{s}$ under
discussion (see later).

The energy of the outgoing photon in the center-of-mass (CM) frame
is given by
%
%
\begin{equation}\label{one}
 E^{CM}_{\gamma}=\frac{s-M_{W}^{2}}{2\sqrt{s}},
\end{equation}
where $s=2m_{e}E_{\bar{\nu}_{e}}$. The range of $\sqrt{s}$ of
interest to us is $\sqrt{s_0} < \sqrt{s} <
\frac{6}{5}\sqrt{s_0}$, where $\sqrt{s_0}\equiv M_W+\Gamma_W $.
 The lower limit for $\sqrt{s}$ is chosen
to distinguish our signature from that due to the soft photon
emission in the resonant absorption process $\bar{\nu}_{e}e^{-}\to
W^-$. The upper limit of $\sqrt{s}$ is the largest incident energy
such that $\bar{\nu}_{e}e^{-}\rightarrow W^{-}\gamma$ remains
dominant over other competing processes (see discussions below).
Substituting $\sqrt{s}=\sqrt{s_{0}}$ in Eq.~(\ref{one}), we obtain
%
%
\begin{equation}\label{two}
 E^{CM}_{\gamma}\simeq \Gamma_{W}.
\end{equation}
We then apply the Lorentz boost to obtain the photon energy in the
Lab frame, or in a high-energy neutrino telescope. For a photon
moving precisely collinear to the incident neutrino, its energy in
the Lab frame is given by
%
%
\begin{equation}\label{three}
 E_{\gamma}\simeq
 \Gamma_{W}\sqrt{\frac{2E_{\bar{\nu}_{e}}}{m_{e}}}\simeq
 \frac{M_W\Gamma_W}{m_e}.
\end{equation}
Note that the boost factor $\sqrt{2E_{\bar{\nu}_e}/m_e}$
significantly enhances the photon energy with respect to its CM
value. According to Eq.~(\ref{three}), $E_{\gamma}$ is roughly
$5\%$ of $E_{\bar{\nu}_{e}}$. We emphasize that this is the
maximal energy that the outgoing photon can carry. Let us also
remark that, for the absorption process
$\bar{\nu}_{e}e^{-}\rightarrow W^{-}\gamma$, the movement of
target electrons inside atoms (in water) does not lead to any
appreciable change in the emitted photon energy.

\begin{figure}[t]
\begin{center}
\leavevmode \epsfxsize=3.5in \epsfysize=3.5in \epsfbox{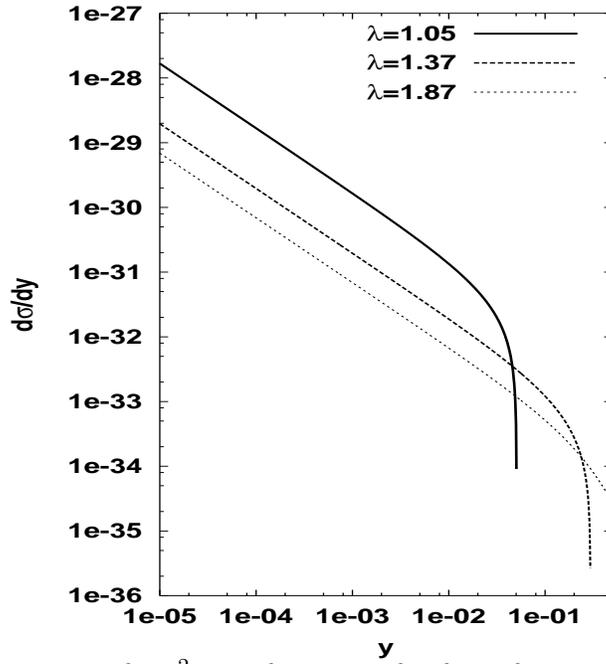}
\tightenlines \caption{The
$\mbox{d}\displaystyle{\sigma}/\mbox{d}\displaystyle{y}$
 in units of cm$^{2}$ as
a function of $y$ for a few representative values of $\lambda $
 [see Eq. (\ref{four})]. $\lambda =1.05$ corresponds to
 $E_{\bar{\nu}_{e}}=6.6\cdot 10^{6}$ GeV, $\lambda =1.37$ corresponds to
 $E_{\bar{\nu}_{e}}=8.6\cdot 10^{6}$ GeV, whereas $\lambda =1.87$
 corresponds to $E_{\bar{\nu}_{e}}=1.1\cdot 10^{7}$ GeV.
 \label{fig:differential}}
\end{center}
\end{figure}

To calculate the differential photon spectrum, it is convenient to
define the dimensionless variables, $\lambda=s/M^{2}_{W}$, and $y=
E_{\gamma}/E_{\bar{\nu}_e}$. In terms of $\lambda$ and $y$, the
differential photon energy spectrum reads:
%
%
\begin{equation}\label{four}
 \frac{\mbox{d}\sigma}{\mbox{d}y}= \frac{\sqrt{2}\alpha
 G_{F}}{\lambda^2(\lambda-1)^2}\displaystyle{
 \left[(\lambda-1)(\lambda^2+1)y^{-1}+4\lambda^2(\lambda-1)y-2\lambda^3 y^2-
 \lambda(3\lambda^2-4\lambda+3)
 \right]}.
\end{equation}
The above equation implies that the final-state photon is more
likely to carry a smaller energy fraction from the incoming
high-energy neutrino, namely it prefers to move back-to-back to
the neutrino. The minimal and maximal values for $y$ are given by
$y_{min}=m_e^2/s$ and $y_{max}=(\lambda-1)/ \lambda$ respectively.
For $\sqrt{s}=\sqrt{s_{0}}$, i.e., $\lambda=1+\Gamma_W/M_W$, we
obtain the value of maximal photon energy as stated in
Eq.~(\ref{three}). Clearly $y_{max}$ increases with the CM energy
$\sqrt{s}\equiv M_W\sqrt{\lambda}$. However, as $\sqrt{s}$
increases, the differential cross section $\mbox{d}\sigma/
\mbox{d}y$ decreases. The behaviors of $\mbox{d}\sigma/ \mbox{d}y$
for a few representative values of $\lambda$ are depicted in Fig.
\ref{fig:differential}. From the behaviors of $\mbox{d}\sigma/
\mbox{d}y$, one can compute the average $y$ value, which is of
more observational interest, as a function of the incoming energy
$\lambda$:
%
%
\begin{equation}\label{five}
\langle y\rangle =\frac{\displaystyle{\int}
 y\frac{\mbox{d}\displaystyle{\sigma}}
 {\mbox{d}\displaystyle{y}}\mbox{d}y}
 {\displaystyle{\int} \frac{\mbox{d}\displaystyle{\sigma}}{\mbox{d}
 \displaystyle{y}}\mbox{d}y}.
\end{equation}
The quantity $\langle y \rangle $ is basically the average
fraction of the incident
 $E_{\bar{\nu}_{e}}$ that is carried by the hard photons in the Lab frame.
The denominator on the R.H.S. of the above equation is just the
total cross
 section
which can be calculated by integrating Eq. (\ref{four}) over $y$.
This gives
%
%
\begin{equation}\label{six}
 \sigma_{\bar{\nu}_{e}e^{-}\rightarrow W^{-}\gamma}=
 \frac{\sqrt{2}\alpha G_{F}}{3\lambda^{2}(\lambda-1)} \left[3(\lambda^{2}+1)\ln
 \left(\frac{M^{2}_{W}(\lambda-1)}{m^{2}_{e}}\right)
       -(5\lambda^{2}-4\lambda+5)\right].
\end{equation}
The cross section $\sigma_{\bar{\nu}_{e}e^{-}\rightarrow W^{-}\gamma}$ given by
Eq. (\ref{six}) depends only on the CM energy $\lambda$.
Numerically, the logarithmic factor in the above equation dominates for
 $\lambda
\simeq 1$ as well as for $\lambda \gg 1$. Our result for
$\sigma_{\bar{\nu}_{e}e^{-}\rightarrow W^{-}\gamma}$ agrees with
the one given by Brown {\em et al.} in Ref. \cite{Brown:1979ux}.
However, it disagrees with the result by Seckel in the same
reference. The numerator on the R.H.S. of Eq.~(\ref{five}) can be
calculated in a similar way. In this case, one may set
$y_{min}=0$. We find
%
%
\begin{equation}\label{seven}
 \displaystyle{\int} y\frac{\mbox{d}\sigma}{\mbox{d}y}dy=\frac{\sqrt{2}\alpha
 G_F}{3\lambda^3} \left(\lambda^2+\lambda+1\right).
\end{equation}
The values of $\langle y \rangle $
 for the energy range of interest to us are as follows:
 $\langle y \rangle
 \simeq 1.3\cdot 10^{-3}$ for
 $E_{\bar{\nu}_{e}}=6.6 \cdot 10^{6}$ GeV, $\langle y \rangle
 \simeq 6.0\cdot 10^{-3}$ for
 $E_{\bar{\nu}_{e}}=8.6 \cdot 10^{6}$ GeV, whereas
 $\langle y \rangle
 \simeq 9.7\cdot 10^{-2}$ for
 $E_{\bar{\nu}_{e}}=1.1 \cdot 10^{7}$ GeV.

\begin{figure}[t]
\begin{center}
\leavevmode \epsfxsize=3.5in \epsfysize=3.5in \epsfbox{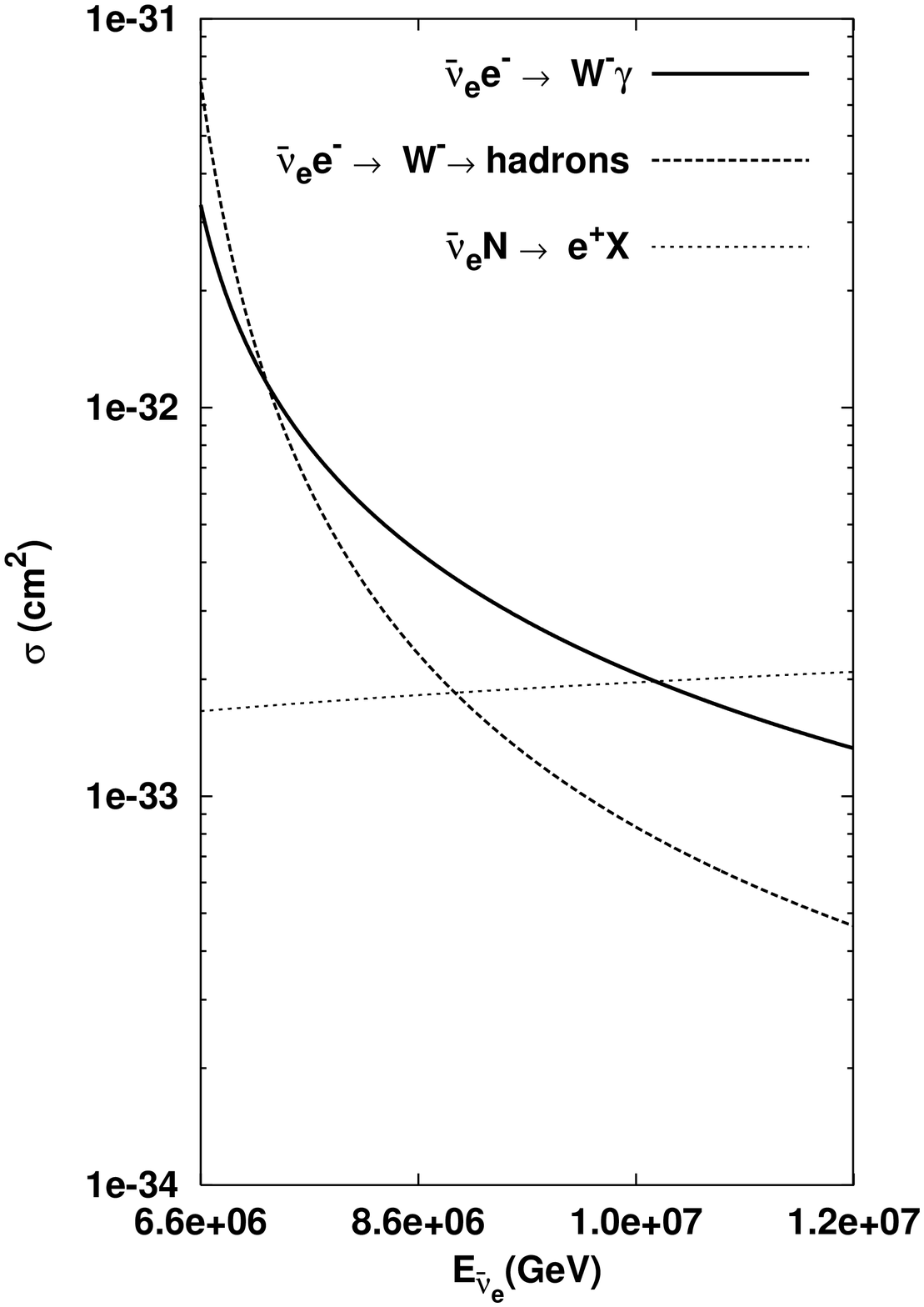}
\tightenlines \caption{High-energy $\bar{\nu}_{e}$ absorption
cross section $\sigma$ (cm$^{2}$), over two different target
particles as a function of electron anti neutrino energy
$E_{\bar{\nu}_{e}}$ (GeV). The minimum value of
$E_{\bar{\nu}_{e}}$ corresponds to
$(M_{W}+\Gamma_{W})^{2}/2m_{e}$. Solid curve is obtained using Eq.
(\ref{six}). Dashed and dotted curves are from Ref.
[5] and Ref. [7], respectively. \label{fig:comparison}}
\end{center}
\end{figure}

It is important to compare the cross section of
$\bar{\nu}_{e}e^{-}\rightarrow W^{-}\gamma$ with  the cross
sections of conventional channels. The behaviors of these cross
sections as functions of $E_{\bar{\nu}_{e}}$ are shown in Fig.
\ref{fig:comparison}.  In this figure, we have included, besides
the cross section of the current process, the resonant cross
section, $\sigma_{\bar{\nu}_{e}e^{-}\rightarrow W^{-}\rightarrow
\mbox{hadrons}}$, taken\footnote{\tightenlines{This cross section
differs from that used in Ref. \cite{Price:1996ep} by a factor of
2. This discrepancy might be due to an incorrect spin averaging
assumed in Ref. \cite{Price:1996ep}.  }} from Ref.
\cite{Gaisser:1995yf}, as well as the charged current deep-inelastic
 $\bar{\nu}_{e}$ scattering cross section over nuclei,
$\sigma^{CC}_{\bar{\nu}_{e}N\rightarrow e^{+}X}$, taken from Ref.
\cite{Gandhi:1998tf} with CTEQ4-DIS parton distributions. We
should remark that the other modern sets of parton distribution
functions only make small differences in the value of
$\sigma^{CC}_{\bar{\nu}_{e}N\rightarrow e^{+}X}$. From Fig.
\ref{fig:comparison}, we note that
$\sigma_{\bar{\nu}_{e}e^{-}\rightarrow W^{-}\gamma}$ dominates
over the other two for  less than half an order of magnitude in
$E_{\bar{\nu}_{e}}$ ($7\cdot 10^{6}\leq
E_{\bar{\nu}_{e}}/\mbox{GeV}\leq 1\cdot 10^{7}$). The dominance is
however within a factor of 2. The effect of this absorption
process can be viewed as an extended (high-energy) tail of the
resonant absorption  process $\bar{\nu}_{e}e^{-}\rightarrow W^{-}$
above the $W$-boson production threshold. This enhancement is due
to the $t$-channel exchange of a nearly on-shell electron [see
Fig. \ref{fig:feynman}(a)]. Note that,  for the relevant
$E_{\bar{\nu}_{e}}$ range, the absorption process
$\bar{\nu}_{e}e^{-}\rightarrow W^{-}\gamma$ is the unique process
with a direct emission of hard photons and a comparable cross
section with the deep-inelastic $\bar{\nu}_{e}$ scattering cross
section over nuclei. Concerning the search for high-energy
$\bar{\nu}_{e}$ , the absorption process over electrons has the
advantage that $\sigma_{\bar{\nu}_{e}e^{-}}$ is basically free
from any theoretical uncertainties unlike
$\sigma_{\bar{\nu}_{e}N}$.

A relevant remark is in order. In Ref. \cite{Athar:2000yw},
assuming an intrinsic relative ratio among high-energy electron,
muon and tau neutrinos as 1: 2: 0 in units of intrinsic electron
neutrino flux, and using the then available constraints on
neutrino mixing parameters (namely $\delta m^{2}$ and
$\sin^{2}2\theta $), it was shown that for cosmologically distant
sources of high-energy neutrinos, this relative ratio becomes 1:
1: 1 again in units of intrinsic electron neutrino flux, due to
neutrino flavor oscillations during the propagation of high-energy
neutrinos. Here electron neutrino stands for the sum of electron
and anti electron neutrino and like wise. In the relevant
$E_{\bar{\nu}_{e}}$ range, the change in the relative ratio of the
intrinsic high-energy neutrino fluxes due to neutrino flavor
oscillations essentially neither depends on $E_{\bar{\nu}_{e}}$,
nor on the value of neutrino mixing parameters, as the three
relative ratios differ by no more than $\sim 10\%$ from 1: 1: 1
(and is due to the present range of uncertainties  in neutrino
mixing parameters). This implies that, in the relevant
$E_{\bar{\nu}_{e}}$ range, the (downward going) intrinsic
high-energy electron neutrino  flux is least affected by neutrino
flavor oscillations, both in its intrinsic energy dependence as
well as in its absolute value. The high-energy muon neutrino flux
changes by $\sim 50\%$, whereas the high-energy tau neutrino flux
changes even more with respect to its intrinsic value. Therefore,
it follows that, for a $E_{\bar{\nu}_{e}}$ range that
$\sigma_{\bar{\nu}_{e}e^{-}\rightarrow W^{-}\gamma}$ dominates
over other competing cross sections, a prospective search for hard
photons in $\bar{\nu}_{e}e^{-}\rightarrow W^{-}\gamma$ may
constrain/measure the {\em intrinsic} $\bar{\nu}_{e}$ flux with
minimal neutrino flavor oscillation effect. In the case where the
constraints for $\nu_{e}$ intrinsic flux exist for the
aforementioned energy range, one can even obtain information on
the neutrino oscillation scenarios that lead to a  change in the
$\bar{\nu}_{e}$ to $\nu_{e}$ ratio with respect to its intrinsic
value (for instance, $\sim 1$) \cite{Mughal:1997xe}.

We have estimated the downward going event rate for the absorption
process $\bar{\nu}_{e}e^{-}\rightarrow W^{-}\gamma $ by using the
following equation:
%
%
\begin{equation}\label{eight}
 \mbox{Rate}= \frac{10}{18}A\int \mbox{d}E_{\bar{\nu}_{e}}P_{\gamma}
 (E_{\bar{\nu}_{e}})\frac{\mbox{d}N}   {\mbox{d}E_{\bar{\nu}_{e}}},
\end{equation}
where $A$ is the area of the high-energy neutrino telescope in ice (or water),
 which we take
 as 1 km$^{2}$, the integration over $E_{\bar{\nu}_e}$
 is performed over the energy range that
 $\bar{\nu}_{e}e^{-}\rightarrow W^{-}\gamma$ dominates,
 and  $P_{\gamma}(E_{\bar{\nu}_{e}})$ is given by

%
%
\begin{equation}\label{nine}
 P_{\gamma}(E_{\bar{\nu}_{e}})=N_{A}\int \mbox{d}y
 R_{\gamma}(y, E_{\bar{\nu}_{e}})
 \frac{\mbox{d}\sigma}
    {\mbox{d}y},
\end{equation}
where $N_{A}=6\cdot 10^{23}\, \mbox{cm}^{-3}$ (water equivalent) is
the Avogadro's number,
$\mbox{d}\displaystyle{\sigma}/ \mbox{d}\displaystyle{y}$ is given
by Eq.~(\ref{four}), while the limits for $y$ are indicated
right after that equation. For simplicity, we take $R_{\gamma}(y,
E_{\bar{\nu}_{e}})\, \simeq R_{e}(y, E_{\bar{\nu}_{e}})$ as an
approximation, where the latter is the electron range given by
 \cite{Gandhi:1996tf,Stanev:1982au}
%
%
\begin{equation}\label{ten}
 R_{e}(y, E_{\bar{\nu}_{e}})\simeq 40 \, \mbox{cmwe}
 \left[(1-\langle y(E_{\bar{\nu}_{e}})\rangle)
 \left(\frac{E_{\bar{\nu}_{e}}}
 {6.2\cdot 10^{4} \, \mbox{GeV}}\right)
 \right]^{\frac{1}{2}}.
\end{equation}
In Eq.~(\ref{eight}), the $\mbox{d}\displaystyle{N}/
\mbox{d}\displaystyle{E_{\bar{\nu}_{e}}}$ is the downward going
differential  $\bar{\nu}_{e}$ flux arriving at the high-energy
neutrino telescope. For $\mbox{d}\displaystyle{N}/
 \mbox{d}\displaystyle{E_{\bar{\nu}_{e}}}$, we consider
the gamma ray burst fireball model proposed in Ref.
\cite{Waxman:1997ti} as an example,
 where $p\gamma $ interactions are suggested
to produce the high-energy $\bar{\nu}_{e}$ flux at the gamma ray burst
 fireball, such that

%
%
\begin{equation}\label{eleven}
 \frac{\mbox{d}N}{\mbox{d}E_{\bar{\nu}_{e}}}\simeq
 \frac{1}{2}\cdot \frac{1}{2} \cdot 4 \cdot 10^{-8}
 \left(\frac{E_{\bar{\nu}_{e}}}{1 \mbox{GeV}}\right)^{-2}
 \mbox{cm}^{-2}\mbox{s}^{-1}\mbox{sr}^{-1}\mbox{GeV}^{-1},
\end{equation}
for the relevant $E_{\bar{\nu}_{e}}$ range. Here we have ignored
the effect of neutrino flavor oscillations in light of earlier
discussions. In Eq.~(\ref{eleven}), the first factor of
$\frac{1}{2}$ arises because half of the electron neutrino flux is
considered to be of anti electron neutrino type and the second
factor of $\frac{1}{2}$ arises because the electron neutrino flux
is considered to be one half of the muon neutrino flux.

The event rate turns out to be $\sim 3\cdot 10^{-4}$ in units of
per year per steradian with 1 km$^{2}$ area and is therefore less than the 
 event rates of other $\bar{\nu}_{e}$ interaction
channels in the same $E_{\bar{\nu}_{e}}$ range
 \cite{Gandhi:1998tf}.
 The event rate for the prospective observation of direct hard photon in the
typical high-energy neutrino telescope is rather low because of
the small $R_{\gamma}$ value, the rather limited range for
$E_{\bar{\nu}_{e}}$, and the high-energy
 $\bar{\nu}_{e}$ flux model being considered. If we take the
 present upper bound for the diffuse high-energy
 neutrino flux set by AMANDA B10 \cite{buda} as the value for
 high-energy $\bar{\nu}_{e}$ flux, then the
 event rate is correspondingly higher up to approximately two orders of
 magnitude.

 Given the current capabilities of typical high-energy neutrino telescopes,
the shower generated by the hard photon is not easy to be
separated from
 the one generated by the $W$-boson inside the high-energy neutrino telescope,
 mainly because of a rather small $R_{\gamma}$ value.
 The signature of this absorption process in terms of event topology
 is thus similar to
that of other $\bar{\nu}_{e}$ interaction channels for the relevant
 $E_{\bar{\nu}_{e}}$ range in the high-energy neutrino telescopes.

In the context of prospective search for high-energy
$\bar{\nu}_{e}$, there can be few other circumstances where the
hard photon emission in the absorption process
$\bar{\nu}_{e}e^{-}\rightarrow W^{-}\gamma$ is of some interest.
 For instance, if a
high-energy $\bar{\nu}_{e}$  crosses a region of
relatively intense $e^{-}$  concentration, the hard
photon with $E_{\gamma}\sim O(10^{4})$ GeV will be emitted along with
the $W$-boson (note, in this context, the absorption process,
$\nu_{e}e^{+}\rightarrow W^{+}\gamma $, also leads to a hard
photon emission along the direction of incident high-energy
$\nu_{e}$ for the same $\sqrt{s}$ range).
 If the scattering length of this hard photon is
less than (or comparable to) the distance between this region and
the prospective detector, one expects that some part of the
 (electromagnetic) shower generated by this hard photon may be
measured. If such a measurement can be implemented, one should be
able to constrain the flux of high-energy $\bar{\nu}_{e}$.
 This type of experiment is complementary to the
prospective direct observation of high-energy $\bar{\nu}_{e}$
 flux through the high-energy neutrino telescopes. An
example can be that the absorption process
$\bar{\nu}_{e}e^{-}\rightarrow W^{-}\gamma $ takes place along the
surface of the earth and the emitted hard photon generates an
 air  shower (as well as the $W$-boson generated shower)
that propagates upward (or in the nearly horizontal direction)
 in the atmosphere of the earth. The future
downward facing space (or possibly balloon) based shower
detectors, if deployed relatively nearby, may eventually become
sensitive to these showers \cite{Domokos:1998hz}. The secondaries
in the nearly horizontal showers generated by hard photons with
 $E_{\gamma}\sim O(10^{4})$ GeV can, in principle, be searched in the
 appropriate ground based detectors as well \cite{Ong:1998ry}.

\acknowledgments H.A. thanks Physics Division of NCTS for
financial support. G.L.L. is supported by the National Science
Council of R.O.C. under the grant number NSC89-2112-M009-041. This
work is carried out under the auspices of the NCTS topical program
{\em Fields and Structures in the Expanding Universe}.

\pagebreak

%
%

%
%

%
%

\begin{references}
\tightenlines
\bibitem{Learned:2000sw}
For recent review articles, see, for instance, J.~G.~Learned and
K.~Mannheim,
Ann.\ Rev.\ Nucl.\ Part.\ Sci.\  {\bf 50}, 679 (2000);
F.~Halzen,
Phys.\ Rept.\  {\bf 333}, 349 (2000), and references cited
therein.
\bibitem{Brown:1979ux}
R.~W.~Brown, D.~Sahdev and K.~O.~Mikaelian,
Phys.\ Rev.\ D {\bf 20}, 1164 (1979);
K.~O.~Mikaelian and I.~M.~Zheleznykh,
Phys.\ Rev.\ D {\bf 22}, 2122 (1980);
V.~S.~Berezinski\u{i} and A.~Z.~Gazizov,
Sov.\ J.\ Nucl.\ Phys.\  {\bf 33}, 120 (1981) [Yad.\ Fiz.\  {\bf
33}, 230 (1981)];
F.~Wilczek,
Phys.\ Rev.\ Lett.\  {\bf 55}, 1252 (1985);
D.~Seckel,
Phys.\ Rev.\ Lett.\  {\bf 80}, 900 (1998) [hep-ph/9709290].
\bibitem{shelly}
S. L. Glashow, Phys.\ Rev. {\bf 118}, 316 (1960).
\bibitem{resonance}
J. N. Bahcall and S. C. Frautschi, Phys.\ Rev. {\bf 135B}, 788
(1960);
V.~S.~Berezinski\u{i} and A.~Z.~Gazizov,
JETP Lett. {\bf 25}, 254 (1977) [Pisma Zh.\ Eksp.\ Teor.\ Fiz.\
{\bf 25}, 276 (1977)];
V.~S.~Berezinsky, D.~Cline and D.~N.~Schramm,
Phys.\ Lett.\ B {\bf 78}, 635 (1978);
I.~M.~Zheleznykh and \'{E}.~A.~Ta\u{i}nov,
 Sov. J. Nucl. Phys. {\bf 32}, 242 (1980)
 [Yad.\ Fiz.\  {\bf 32}, 468 (1980)];
V.~S.~Berezinsky and V. L. Ginzburg, Mon. Not. R. Astr. Soc. {\bf
194}, 3 (1981);
R.~W.~Brown and F.~W.~Stecker,
Phys.\ Rev.\ D {\bf 26}, 373 (1982);
G.~Domokos and S.~Kovesi-Domokos,
Phys.\ Lett.\ B {\bf 346}, 317 (1995) [hep-ph/9410352].
\bibitem{Gaisser:1995yf}
T.~K.~Gaisser, F.~Halzen and T.~Stanev,
Phys.\ Rept.\  {\bf 258}, 173 (1995) [Erratum-ibid.\  {\bf 271},
355 (1996)] [hep-ph/9410384].
\bibitem{Price:1996ep}
P.~B.~Price,
Astropart.\ Phys.\  {\bf 5}, 43 (1996)
[astro-ph/9510119].
\bibitem{Gandhi:1998tf}
R.~Gandhi, C.~Quigg, M.~H.~Reno and I.~Sarcevic,
Phys.\ Rev.\ D {\bf 58}, 093009 (1998) [hep-ph/9807264].
\bibitem{Athar:2000yw}
H.~Athar, M.~Je\.{z}abek and O.~Yasuda,
Phys.\ Rev.\ D {\bf 62}, 103007 (2000) [hep-ph/0005104], and
references cited therein.
\bibitem{Mughal:1997xe}
See, for instance, M.~A.~Mughal and H.~Athar,
hep-ph/9806408.
\bibitem{Gandhi:1996tf}
R.~Gandhi, C.~Quigg, M.~H.~Reno and I.~Sarcevic,
Astropart.\ Phys.\  {\bf 5}, 81 (1996)
[hep-ph/9512364].
\bibitem{Stanev:1982au}
 T.~Stanev, C.~Vankov, R.~E.~Streitmatter, R.~W.~Ellsworth and T.~Bowen,
Phys.\ Rev.\ D {\bf 25}, 1291 (1982).
\bibitem{Waxman:1997ti}
E.~Waxman and J.~Bahcall,
Phys.\ Rev.\ Lett.\  {\bf 78}, 2292 (1997)
[astro-ph/9701231].
\bibitem{buda} See, for instance, M. Kowalski, talk given at
 {\em International Europhysics Conference on High Energy Physics}
 (EPS HEP 2001), Budapest (Hungary), 2001 (to appear in its proceedings).
\bibitem{Domokos:1998hz}
G.~Domokos and S.~Kovesi-Domokos,
hep-ph/9805221;
D.~Fargion,
astro-ph/0002453.
For a recent review article, see, D.~B.~Cline and F.~W.~Stecker,
astro-ph/0003459.
\bibitem{Ong:1998ry}
 See, for instance, R.~A.~Ong,
Phys.\ Rept.\  {\bf 305}, 93 (1998).
\end{references}
\end{document}